\documentclass{JHEP3}
\usepackage{amsmath,amsthm,latexsym}
\newcommand{\be}{\begin{equation}}
\newcommand{\ee}{\end{equation}}
\preprint{{\tt hep-th/yymmnnn}}

\title{ \center{Analyticity Properties of Graham-Witten Anomalies}}

\author{
Vadim Asnin\footnotemark[1]\\
 Racah Institute of Physics\\
 Hebrew University \\
 Jerusalem 91904,
 Israel\\

 \footnotemark[1] {\tt vadim.asnin@mail.huji.ac.il}}

\abstract{Analytic properties of Graham-Witten anomalies are
considered. Weyl anomalies according to their analytic properties
are of type A (coming from $\delta$-singularities in correlators
of several energy-momentum tensors) or of type B (originating in
counterterms which depend logarithmically on a mass scale). It is
argued that all Graham-Witten anomalies can be divided into 2
groups: internal and external, and that all external anomalies are
of type B, whereas among internal anomalies there is one term of
type A and all the rest are of type B. This argument is checked
explicitly for the case of a free scalar field in a 6-dimensional
space with a 2-dimensional submanifold.}

\keywords{Graham-Witten anomaly, Weyl anomaly}
\begin{document}

\section{Introduction}
  ~~~~From the beginning of 1970's we know that dilation and conformal invariance of quantum field
  theories are in general broken by quantum effects. There are two different kinds of
  breaking:
  1) The equations of motion for operators do not coincide with classical ones, i.e. the theory
  has a nonvanishing $\beta$-function;
  2) The $\beta$-function is 0, so the theory is conformally invariant also at the quantum level, but in certain
  correlators the corresponding Noether currents  possess anomalous
  divergences. This phenomenon was called ``trace (Weyl, conformal)
  anomaly". (The first case appears in the theories like QED or QCD. It is described, for example, in
  \cite{Wilson:1973jj,Callan:1970yg,Callan:1970ze}. In what follows we deal only with the second case described in
  \cite{Deser:1976yx}).

  The problem of identifying and classifying all possible trace anomalies was extensively studied. In
  \cite{Bonora:1983ff,Bonora:1985cq} this problem was treated by the grading of Weyl variation operator and it was shown
  that the trace anomalies appear due to a nontrivial cohomology of the Weyl transformations
  in the space of local diffeoinvariant polynomials. This approach naturally explains why, for example, one
  sometimes can get rid of an anomaly by changing the procedure of regularizing the ultraviolet divergences,
  and allows one to identify on general grounds the anomalies that can be disposed of in this way.

  An approach to the classification of the trace anomalies taking into account the mechanism producing them
  was suggested in \cite{Deser:1993yx}. It was shown that all local trace anomalies can be divided into two classes
  (types A and B) with different production mechanisms. Anomalies of type A appear because of the
  $\delta$-singularities in the discontinuities of correlators of several energy-momentum tensors;
  the anomaly itself is always equal to the Euler density. On the other hand, anomalies of type B
  arise because of counterterms that must be added to the effective action in order to make finite
  certain logarithmically divergent correlators. The corresponding anomalies are given by Weyl invariant polynomials.

  The way in which the trace anomalies are reproduced by the gravitational
  action in the framework of the $AdS/CFT$ duality was studied in \cite{Henningson:1998gx}. In this correspondence one can compute
  the generating functional of the conformal field theory on some $D$-dimensional manifold $M$ by
  evaluating the gravitational action for the metric on some $D+1$-dimensional manifold $X$ which has
  $M$  as its conformal boundary. This gravitational action turns out to have (infrared) divergences due to the
  integration over the whole $D+1$-dimensional volume. Covariant regularization of the divergence breaks the
  Weyl invariance and an anomaly is produced. This is consistent with the fact that conformal field theory on
  $M$ also has a Weyl anomaly, and this anomaly is reproduced by the anomaly of the regularized gravitational
  action of $X$.

  This idea can be generalized in a following sense: expectation values of observables which are defined on
  a submanifold can be anomalous because of the integration over the ambient space. This anomaly (we will call
  it Graham-Witten, or GW anomaly \cite{Graham:1999pm}) is more general than previously known ones since it deals with
  submanifolds in Riemann or Pseudo-Riemann spaces,  whereas usually one considers these spaces themselves. If the
  submanifold coincides with the whole space then we are in the standard situation. Other cases must be studied
  separately. In a general situation new terms appear in the expression of the anomaly. These terms are due
  to the presence of the ambient space.

  The GW anomaly can be studied independently of the $AdS/CFT$ correspondence. This anomaly  was calculated
  directly in the framework of the quantum field theory for various models in
  \cite{Henningson:1999xi,Gustavsson:2004gj}.  However, the methods
  used in the calculations do not reveal the mechanism of the anomaly. In this paper we fulfill this gap by
  presenting a different method which
  is in close relation to that of \cite{Deser:1993yx}. This method allows one to understand an origin of various terms
  in the expression of the anomaly. The method was developed in
  \cite{Master}.

  The paper is organized in a following way. Part 2 describes the classification of Weyl anomalies.
  Part 3 contains results concerning the GW anomaly, especially those obtained in \cite{Henningson:1999xi}.
  In part 4 we consider the general properties of the GW anomaly and present the calculation for massless scalar field
  in 6 dimensions (this model was treated in a different way in \cite{Gustavsson:2004gj}).
  On this model we determine the production mechanisms of various terms in the expression of the anomaly.
  Part 5 is the summary of the results.


\section{Classification of Weyl anomalies}
  ~~~~A classification of the Weyl anomalies according to the mechanism of their appearance was given in \cite{Deser:1993yx}.
  It is based on the observation that the integrated trace anomaly can be considered as a variation
  of an effective action w.r.t. the mass scale. Then if this integrated anomaly vanishes then the effective
  action contains no mass scale. This type of the trace anomaly was called in \cite{Deser:1993yx} ``type
  A". If, on the other hand, the integrated anomaly doesn't vanish then the effective action contains a mass scale
  and the corresponding anomaly is called ``type B".

  As observed in \cite{Deser:1993yx} the distinction between the 2 types of anomalies is clear if one uses the
  dimensional regularization for treating the divergences.

  Consider the conformal field theory on a curved $D$-dimensional Riemann manifold $M$ with some metric $g_{\mu\nu}$. We
  introduce the effective action $W[g_{\mu\nu}]$ defined as \be e^{i
  W[g_{\mu\nu}]}=\int\mathcal{D}
  \phi\, e^{iS[g_{\mu\nu},\phi]}.\ee In order to regularize
  the ultraviolet divergences one takes $D$ to be non-integer but close to some even number $2n$ (we know that there
  are no anomalies in odd dimensions). Then $W[g_{\mu\nu}]$ becomes finite and no anomalies can appear. However,
  $W[g_{\mu\nu}]$ can have a pole term at $D=2n$, meaning that it looks like $W'[g_{\mu\nu}]/\epsilon$, where
  $\epsilon=D-2n$. This $1/\epsilon$ term is essentially the source of the anomaly in this approach. There are 2
  possibilities: $W'[g_{\mu\nu}]$ may or may not vanish at $D=2n$. If $W'[g_{\mu\nu}]$ vanishes then the type A anomaly
  arises, otherwise the corresponding anomaly is type B.

  The typical example of the type A situation is the trace anomaly in 2 dimensions which is proportional to the curvature of the space.
  It is shown in \cite{Deser:1993yx} that the effective action for $D\approx 2$ is
    \be W_D[g_{\mu\nu}]=-\frac14\int\limits_M d^D x\sqrt{g}\,R\,\Box^{-1-\epsilon/2}R\,+\,\frac{1}{\epsilon}\int
    \limits_M d^D\label{WD}x\sqrt{g}\,R. \ee
  Here $M$ is the $D$-dimensional manifold which is obtained from the initial 2-dimensional Riemann space by some kind
  of an ``analytic continuation," and $g_{\mu\nu}$ and $R$ denote the metric and the curvature of this continued manifold.
  This effective action is Weyl invariant to the order $\epsilon$ in $D$ dimensions.

  The second term in this expression has $0/0$ ambiguity since the integral vanishes at $D=2$. To take the limit is
  not trivial, since it depends on how $M$ was continued to $D$ dimensions. One can imagine, for example, that the
  integral in the second term vanishes also for the continued manifold. Then the second term disappears and we are left
  with a finite expression in 2 dimensions (Polyakov action)
    \be W_2[g_{\mu\nu}]=-\frac14\int\limits_M d^2 x\sqrt{g}\,R\,\Box^{-1}R.\label{WD2} \ee
  We see that this way of taking the limit is just a kind of regularization of the effective action.

  To compute the Weyl variation of $W$ one just notices that in $D$ dimensions $\delta W_D=0$, so the Weyl variations
  of two terms in eq.(\ref{WD}) cancel. Therefore the Weyl variation of $W_2$ as we defined it is opposite in sign to that
  of the dropped local term in (\ref{WD}). The latter can be calculated easily, giving
    \be \delta W_2\sim\int\limits_M d^2x\sqrt{g}\,R\,\sigma.\label{R2DS} \ee
  In the final answer the $1/\epsilon$ cancels, and the result is finite. It is the general feature of the trace
  anomalies. It was shown in \cite{Deser:1993yx} that similar anomalies of type A appear in any
  even dimension and the corresponding anomaly density is the Euler density.

 The simplest example of type B anomaly appears in four dimensions (the so-called $c$-anomaly).
  As it was shown in \cite{Deser:1993yx} the corresponding effective action is
    \be W_D[g_{\mu\nu}]=\frac{1}{\epsilon}\int\limits_Md^Dx\sqrt{g}\,C_{\mu\nu\xi\eta}
    \,\Box^{\epsilon/2}C^{\mu\nu\xi\eta}, \label{W2} \ee
  where $C_{\mu\nu\xi\eta}$ is the Weyl tensor which is like all the other continued to $D$ dimensions. $W_D$ is again
  Weyl invariant to order $\epsilon$ but it diverges at $D=4$, so one should modify it by adding a suitable counterterm
  that would produce a finite result. The subtracted $W_D$ is
    \be
    W_D^{sub}[g_{\mu\nu}]=W_D[g_{\mu\nu}]-\frac{\mu^\epsilon}{\epsilon}\int\limits_Md^Dx\sqrt{g}\,
       C_{\mu\nu\xi\eta}C^{\mu\nu\xi\eta},\label{W2S} \ee
  where $\mu$ is some arbitrary mass. Then $W_4$ can be rewritten as
    \be W_4[g_{\mu\nu}]=\frac12\int\limits_Md^Dx\sqrt{g}\,C_{\mu\nu\xi\eta}\log\Bigl(\frac{\Box}{\mu^2}\Bigr)
       C^{\mu\nu\xi\eta}+\frac{\mu^\epsilon}{\epsilon}\int\limits_Md^Dx\sqrt{g}\,
       C_{\mu\nu\xi\eta}C^{\mu\nu\xi\eta},\label{W2Sfull} \ee
  As in the case of the type A anomaly, the effective action is a sum of nonlocal finite term and a pole term. However, the residue in this case
  doesn't vanish.
  One of the ways to take the limit $\epsilon\rightarrow0$ is just to drop the pole term:
    \be W_4[g_{\mu\nu}]=\frac12\int\limits_Md^Dx\sqrt{g}\,C_{\mu\nu\xi\eta}\log\Bigl(\frac{\Box}{\mu^2}\Bigr)
       C^{\mu\nu\xi\eta}.\label{WeylTensor} \ee
  Again, the Weyl variation of the nonlocal term is opposite in sign to that of the pole term, so
    \be \delta W_4[g_{\mu\nu}]\sim\int\limits_Md^4x\sqrt{g}\,C_{\mu\nu\xi\eta}C^{\mu\nu\xi\eta}\,\sigma, \ee
  which is finite like in the case of type A.

  As stated in the beginning of this section, the essential difference between the 2 types is that the anomalies
  of type B emerge as a consequence of the mass
  scale $\mu$ that is introduced to the theory by counterterms. On the other hand, the anomalies of type A arise without
  the scale and reflect the reduction in the number of Weyl invariant expressions when one crosses an even dimension.

  In part 4 we will use this classification in order to identify the types of GW anomalies.


\section{The Graham-Witten Anomaly}
  ~~~~In this part we present the results obtained mostly in \cite{Henningson:1999xi}. In this paper the GW anomaly was calculated for
  the free self-dual gauge field on the 6-dimensional Riemann manifold. Here the submanifold for which the GW anomaly
  should be calculated is just the 2-dimensional surface of the Wilson observable. This observable is dimensionless and,
  hence, Weyl invariant classically.

  So, consider the 6-dimensional Riemann space $M$ and the 2-dimensional submanifold $N$ embedded into $M$. We denote the
  coordinates in $M$ as $X^{\mu},\:\mu=1...6$ and the coordinates in $N$ as $\xi^{\alpha},\:\alpha=1...2$. The metric on
  $M$ is denoted as $G_{\mu\nu}$. The corresponding induced metric on $N$ is $g_{\alpha\beta}$. We also denote as
  $\nabla_{\mu}$ and $\hat\nabla_{\alpha}$ the covariant derivatives in the ambient space along the coordinate line
  $\mu$ and on the submanifold along the coordinate line $\alpha$ correspondingly. We will also use the Christoffel
  symbols of $M$ and $N$ which we denote as $\Gamma_{\mu\nu}^{\lambda}$ and $\hat\Gamma_{\alpha\beta}^{\gamma}$. The
  Riemann tensors, the Ricci tensors and the curvatures of $M$ and $N$ we denote as $R_{\mu\nu\xi\eta},\,
  \hat R_{\alpha\beta\gamma\delta},\, R_{\mu\nu},\,\hat R_{\alpha\beta},\,R$ and $\hat R$.

  The gauge field in 6 dimensions is the antisymmetric tensor $A_{\mu\nu}$. The action of the free gauge field is
    \be S=-\frac{1}{12}\int\limits_Md^{\,6}X\sqrt{G}\,F_{\mu\nu\lambda}F^{\mu\nu\lambda}, \ee
  where $F_{\mu\nu\lambda}=3\nabla_{[\lambda}A_{\mu\nu]}$ is the field strength. The action is invariant under the gauge
  transformations $A_{\mu\nu}\rightarrow A_{\mu\nu}+\overline A_{\mu\nu}$, where $\overline A_{\mu\nu}$ is exact form with
  integer periods.

  In what follows we will need the propagator in the Feynmann gauge of this field which we denote as
  $\Delta_{\mu\nu,\xi\eta}$:
    \be \Delta_{\mu\nu,\xi\eta}(X,Y)=<0|A_{\mu\nu}(X)A_{\xi\eta}(Y)|0>. \ee
  It is an antisymmetric bitensor.

  Given the 2-dimensional submanifold $N$, one can construct the Wilson surface observable
    $$W[N]=\exp\Bigl(2\pi i\int\limits_N\,d\xi^{\alpha}\wedge d\xi^{\beta}\,\tilde A_{\alpha\beta}\Bigr),
    \label{WilsonLoop}$$
  $\tilde A_{\alpha\beta}=\partial_{\alpha}X^{\mu} \partial_{\beta}X^{\nu}A_{\mu\nu}$ being the pull-back of $A_{\mu\nu}$
  to $N$. The vacuum expectation value of $W$ is
    $$<0|W[N]|0>=e^{-4\pi^2I},$$
  with
    \be I=\int\limits_Nd\xi_1^{\alpha}\wedge d\xi_1^{\beta}\int\limits_Nd\xi_2^{\gamma}\wedge
       d\xi_2^{\delta}\,\tilde{\Delta}_{\alpha\beta,\gamma\delta}(\xi_1,\xi_2),\label{I} \ee
  where
    $$\tilde{\Delta}_{\alpha\beta,\gamma\delta}(\xi_1,\xi_2)=\partial_{\alpha}X^{\mu}(\xi_1)\partial_{\beta}X^{\nu}(\xi_1)
    \partial_{\gamma}X^{\xi}(\xi_2)\partial_{\delta}X^{\eta}(\xi_2)\,\Delta_{\mu\nu,\xi\eta}(X,Y)$$
  is the pull-back of the propagator to the submanifold $N$. Formally one can say that $<0|W[N]|0>$ is still conformally
  invariant even after adding the gauge-fixing term to the Lagrangian. But it's not true, however, since $I$ diverges,
  and the regularization leads to the anomaly.

  There are many ways to regularize $I$ without breaking the diffeomorphism invariance. For example, $N$ can be
  replaced by two 2-dimensional surfaces separated by an infinitesimal distance $\delta$ (this regularization is used in \cite{Gustavsson:2004gj}).
  In the regularization used
  in \cite{Henningson:1999xi} the points $\xi_1$ and $\xi_2$ on $N$ are prevented form being closer to each other then some
  infinitesimal $\epsilon$. To preserve the diffeomorphism invariance $\epsilon$ should be measured by the geodesic
  distance.

  If one adopts this last regularization scheme then, as it was shown in \cite{Henningson:1999xi}, the conformal variation of $I$ is
    \begin{multline}
       \delta I=\int\limits_Nd\xi^1d\xi^2\sqrt{g}\Bigl[\frac{4}{\epsilon^2}\sigma-\frac12\hat
       R\sigma-\frac34\Bigl((\Box X)^2-4g^{\alpha\beta}\tilde P_{\alpha\beta}\Bigr)\sigma-\\
       -\frac16g^{\alpha\gamma}g^{\beta\delta}\tilde C_{\alpha\beta\gamma\delta}\,\sigma- \frac56{\Box}\label{HSMain}
       X^{\mu}\nabla_{\mu}\sigma\Bigr].
    \end{multline}
  Here $\sigma$ is a parameter of the infinitesimal Weyl transformation, $\tilde C_{\alpha\beta\gamma\delta}$ is
  a pull-back of the Weyl tensor of the ambient space $C_{\mu\nu\xi\eta}$, $\tilde P_{\alpha\beta}$ is a pull-back of
  a tensor $P_{\mu\nu}$ which is defined as
    \be P_{\mu\nu}=\frac14(R_{\mu\nu}-\frac{1}{10}RG_{\mu\nu}),\label{Pmunu} \ee
  $\Box=\hat\nabla_{\alpha}\hat\nabla^{\alpha}$ is a Laplacian on $N$ and $\Box X^{\mu}$ is a mean curvature
  vector which is a trace of a second fundamental form $\Omega_{\alpha\beta}^{\,\mu}$:
    \be \Omega_{\alpha\beta}^{\,\mu}=\partial_{\alpha}\partial_{\beta}X^{\mu}-\hat\Gamma_{\alpha\beta}^{\delta}
       \partial_{\delta}X^{\mu}+\Gamma_{\nu\lambda}^{\mu}\partial_{\alpha}X^{\nu}\partial_{\beta}X^{\lambda}\label{SFF} \ee
  We see that the conformal variation diverges in the limit $\epsilon\rightarrow0$. However, one can consider
  the renormalized Wilson observable
    \be W_R=W\int\limits_{N}d\xi^1d\xi^2\Bigl(-\frac{2}{\epsilon^2}\sqrt{g}\Bigr).\ee
  The conformal variation of $W_R$ is finite. It is given by
    \begin{multline}
       \delta W_R=-4\pi^2\,W_R\,\int\limits_Nd\xi^1d\xi^2\sqrt{g}\Bigl[-\frac12\,\hat R\,\sigma-\frac34\Bigl((\Box X)^2-
       4g^{\alpha\beta}\tilde P_{\alpha\beta}\Bigr)\,\sigma-\\
       -\frac16\,g^{\alpha\gamma}g^{\beta\delta}\tilde C_{\alpha\beta\gamma\delta}\,\sigma-\frac56\,{\Box}X^{\mu}
       \nabla_{\mu}\sigma\Bigr].\label{HSResult}
    \end{multline}
  This expression agrees with the general features of the trace anomaly mentioned
  above. However, as noticed in \cite{Schwimmer:2008yh} the
  last term represents a trivial anomaly since it is a Weyl
  variation of a local expression:$$\int\limits_{N}d\xi^1d\xi^2\sqrt{g}\Box
  X^{\mu}\nabla_{\mu}\sigma\propto\delta\int\limits_{N}d\xi^1d\xi^2\sqrt{g}\Box
  X^{\mu}\Box X_{\mu},$$ and therefore the true anomaly is given
  by just 3 first terms.

  In the next part we are going to examine this result in detail and, in particular, answer the following question:
  What type is this anomaly?


\section{Analysis of GW anomaly}
  ~~~~In this part we present a general analysis of the GW anomalies and give
  a detailed calculation in the simplest case of 2-dimensional
  surface embedded into a 6-dimensional space.

  It is well-known that the trace anomaly can appear only in spaces of
  even dimension.  It follows, for example, from the cohomological
  analysis of \cite{Bonora:1983ff, Bonora:1985cq}. In the case of the GW anomaly the
  precise connection between the dimension $D$ of the ambient space
  and the dimension $d$ of the surface of an observable is
      \be d=\frac{D-2}{2}\label{dD} \ee
  For odd-dimensional spaces (odd $D$) the dimensions of the ``conformally invariant" submanifolds is to be half-integer,
  which is impossible. This result essentially shows that for odd $D$ one cannot construct the self-dual gauge field theory.
  Such a field is a scalar in 2 dimensions, a vector in 4 dimensions, an antisymmetric tensor of the rank 2 in 6 dimensions
  etc. There is no room for odd dimensions in this scheme. The same situation occurs if we try to construct nonlocal
  variables from scalars as we'll discuss in the following.

  In order to proceed we will need to make a few general observations about the conformal variations on submanifolds.
   Suppose that we make the conformal variation of the metric in the ambient space $M$. The change in the metric is
    \be \delta G_{\mu\nu}=2\sigma G_{\mu\nu}\label{ConfVar}. \ee
  Together with this the induced metric on the submanifold $N$
  $g_{\alpha\beta}=\partial_{\alpha}X^{\mu}\partial_{\beta}X^{\nu}
  G_{\mu\nu}$ undergoes changes as well. Since the derivatives of the coordinates are invariant under the conformal
  transformations, $g_{\alpha\beta}$ under (\ref{ConfVar}) changes in way similar to $G_{\mu\nu}$
  \be \delta g_{\alpha\beta}=2\sigma g_{\alpha\beta}\label{ConfVarN}.\ee Therefore all the combinations of
  $g_{\alpha\beta}$ and its derivatives, $\hat R_{\mu\nu\xi\eta}$ for example, change
  under the transformation as if there were no ambient space at all. Weyl anomalies are local diffeoinvariant
  expressions constructed of ``geometrical objects" of a definite dimension and linear in $\sigma$.
  Therefore we can say that among all possible GW anomalies there are always those that ``belong" just
  to the submanifold $N$. These are the usual trace anomalies in the space of dimension $d$.
  Notice that they exist only for even $d$.

  If there were no ambient space $M$ then these would be the only possible anomalies. But the presence of $M$ allows
  for new local diffeoinvariant expressions which are built from the terms among which there are some that don't ``belong"
  to $N$. We will call the anomalies that belong to $N$ ``internal" and all other ``external".

  Consider now the local terms
  that don't belong to $N$. They can involve, for example, the Riemann tensor, the Ricci tensor and the curvature of $M$,
  the  second fundamental form and their covariant derivatives. One can easily check that any such expression with all the
  indices contracted is of even mass dimension. Therefore the GW anomaly apparently is possible only if $N$
  is even dimensional. For the free self-dual gauge field it can occur according to (\ref{dD}) only if
    \be D=2+4n,\qquad n=0,1,2... \ee
  The case $D=2$ is trivial since $N$ is of dimension $0$ and is just a set of points. The first non-trivial case is
  $D=6$ which is dealt with in \cite{Henningson:1999xi}.

  We are now going to show that all external anomalies are of type B. We saw in part 2 that the anomalies of type B
  arise as conformal variations of pole terms with non-vanishing residue. What we will argue is that any pole term
  that produces an anomaly generically doesn't vanish. To do so notice first that no external anomaly can be a conformal
  variation of an internal expression since any such variation is also internal.
  Then suppose that some external anomaly is a conformal variation of a pole term with vanishing residue. This residue
  is an integral over $N$ of an expression which contains some ``external" quantities like the second fundamental form,
  for example. Then we can change $M$ at least in some small domain in such a way that $N$ (and therefore all internal quantities)
  will remain the same but some external quantities will change.
  Then the value of the integral will change. So we see that the residue doesn't generically vanish.

  Similar situation occurs with the usual type B anomalies. Consider again the example of type B anomaly given in part 2.
  One can imagine that for some particular manifold the integral in the second term in the eq. (\ref{W2Sfull}) vanishes.
  But it doesn't mean that the corresponding anomaly is of type A since we are to consider generic manifolds.

  Consider the case $D=6$ and $d=2$ in more detail. In order to do this we instead of $\delta I$ as was done in
  \cite{Henningson:1999xi,Gustavsson:2004gj} (see eq. (\ref{HSMain})) calculate $I$ itself (the
  calculation was carried out in \cite{Master}). It turns out that for our purposes one
  doesn't really have to work with the gauge field. This is so because the
  only essential features of the theory that are necessary for us are its symmetries. There are 3 different symmetries
  here: 1) symmetry with respect to the diffeomorphisms of $M$, 2) symmetry with respect to the diffeomorphisms of $N$,
  3) Weyl symmetry in $M$ (and therefore in $N$).
  Presence of the GW anomaly tells us that no regularization can preserve all of them, and this fact is independent of any
  particular theory. Therefore we can choose any theory we like and do all the calculations with it. We choose the simplest
  possible theory, the free massless scalar field theory. This choice was suggested in
  \cite{Gustavsson:2004gj, Master}. But instead of the point-splitting regularization of
  \cite{Henningson:1999xi} and \cite{Gustavsson:2004gj} we implement the dimensional
  regularization for the reasons explained above.

  The action of free massless scalar field $\phi$ in $D$-dimensional curved space $M$ is given by
    \be S=\int\limits_Md^Dx\sqrt{G}\,(G^{\mu\nu}\partial_{\mu}\phi\,\partial_{\nu}\phi-AR\,\phi^2), \ee
  where
    \be A=\frac{D-2}{4(D-1)}. \ee
  With the usual Weyl transformation law for the scalar field $\delta\phi=-\frac{D-2}{2}\sigma\phi$ this action is
  Weyl invariant (our convention for the curvature is $R_{\mu\rho\nu\sigma}=
  G_{\sigma\xi}(\partial_{\mu}\Gamma_{\rho\nu}^{\xi}-\partial_{\rho}\Gamma_{\mu\nu}^{\xi})+...$).

  For any $\frac{D-2}{2}$ - dimensional submanifold $N$ we, following \cite{Gustavsson:2004gj}, construct the analog of the Wilson loop
    \be W[N]=\exp\Bigl(2\pi i\int\limits_Nd^{\frac{D-2}{2}}\xi\sqrt{g(\xi)}\,\phi(\xi)\Bigr). \ee
  Then we can evaluate its vacuum expectation value
    \be  <0|W[N]|0>=\int\mathcal{D}\phi\,W[N]e^{-S}.\ee
  This path integral can be calculated exactly and the result is
    \be <0|W[N]|0>=e^{-4\pi^2I}, \ee
  where $I$ is given now by
    \be
    I=\int\limits_Nd^{\frac{D-2}{2}}\xi_1\sqrt{g(\xi_1)}\int\limits_Nd^{\frac{D-2}{2}}\xi_2\sqrt{g(\xi_2)}\,\Delta(\xi_1,\xi_2).\label{I1} \ee
  $\Delta$ here is the propagator of the scalar field. Consider 2
  nearby points in $M$, which we denote by $X_0$ and $X$. We
  introduce the Riemann normal coordinates $X^{\mu}$ on $M$ at $X_0$. In these coordinates the
  short-distance expansion of the propagator between $X_0$ and $X$ is
  \be
  \Delta(X,X_0)=\frac{\alpha}{|X|^{D-2}}\Bigl(1-\frac{D-2}{12}P_{\mu\nu}X^{\mu}X^{\nu}+\:...\:\Bigr).\label{Prop}\ee
  Here \be
  P_{\mu\nu}=\frac{1}{D-2}\Bigl(R_{\mu\nu}-\frac{R}{2(D-1)}G_{\mu\nu}\Bigr),\ee
  where $R_{\mu\nu}$ and $R$ are the Ricci tensor and the curvature
  scalar of $M$ at $X_0$ (for $D=6$ it coincides with the tensor defined in (\ref{Pmunu})).
  This tensor possesses a very simple  conformal variation \be \delta
  P_{\mu\nu}=\nabla_{\mu}\nabla_{\nu}\sigma.\ee The coefficient
  $\alpha$ in (\ref{Prop}) is given by \be
  \alpha=-\frac{1}{(D-2)\,\sigma_D},\label{al}\ee with
  $\sigma_D=2\pi^{D/2}/\Gamma(\frac D2)$ being an area of a unit sphere
  in $D$-dimensional flat Euclidean space.

  In order to calculate the propagator between 2 points on the submanifold $N$ we need the expression for $|X|$ in the
  coordinates $\xi$ on $N$. For this purpose consider 2 nearby points $A_0$ and $A$ on $N$. At $A_0$ we introduce
  normal coordinates on $N$ that we denote by $u^{\alpha}$ (in addition to the previously introduced normal coordinates in $M$).
  We consider the difference of the $X$-coordinates of $A$ and $A_0$:
    \be  X^{\mu}(A)=X^{\mu}(A_0)+\partial_{\alpha}X^{\mu}(A_0)\,u^{\alpha}+\frac12\partial_{\alpha}\partial_{\beta}
       X^{\mu}(A_0)\,u^{\alpha}u^{\beta}
       +\frac16\partial_{\alpha}\partial_{\beta}\partial_{\gamma}X^{\mu}(A_0)\,u^{\alpha}u^{\beta}u^{\gamma}+\:...\ee
  For $|X(A)-X(A_0)|^2$ we have
    \begin{multline*}
       |X(A)-X(A_0)|^2=\partial_{\alpha}X^{\mu}\partial_{\beta}X_{\mu}u^{\alpha}u^{\beta}+\partial_{\alpha}
       \partial_{\beta}X^{\mu}\partial_{\gamma}X_{\mu}u^{\alpha}u^{\beta}u^{\gamma}+\\
       +\Bigl(\frac14\partial_{\alpha}\partial_{\beta}X^{\mu}\partial_{\gamma}\partial_{\delta}X_{\mu}+\frac13\partial_
       {\alpha}\partial_{\beta}\partial_{\gamma}X^{\mu}\partial_{\delta}X_{\mu}\Bigr)u^{\alpha}u^{\beta}u^{\gamma}
       u^{\delta}+\:...\hspace{1cm}
    \end{multline*}
  In the first term $\partial_{\alpha}X^{\mu}\partial_{\beta}X_{\mu}$ is the induced metric at $A_0$ which is
  $\delta_{\alpha\beta}$. The second term vanishes because in the normal coordinates $\partial_{\alpha}\partial_{\beta}
  X^{\mu}$ is orthogonal to $N$. We use this last fact to get
    \be
       \partial_{\alpha}\partial_{\beta}\partial_{\gamma}X^{\mu}\partial_{\delta}X_{\mu}=\partial_{\alpha}\Bigl(
       \partial_{\beta}\partial_{\gamma}X^{\mu}\partial_{\delta}X_{\mu}\Bigr)-\partial_{\alpha}\partial_{\beta}
       X^{\mu}\partial_{\gamma}\partial_{\delta}X_{\mu}
       =-\partial_{\alpha}\partial_{\beta}X^{\mu}\partial_{\gamma}\partial_{\delta}X_{\mu}
    \ee
  and then for $|X(A)-X(A_0)|^2$ we obtain
    \be |X(A)-X(A_0)|^2=u^2-\frac{1}{12}\delta_{\mu\nu}\Omega_{\alpha\beta}^{\,\mu}\Omega_{\gamma\delta}^{\,\nu}
       u^{\alpha}u^{\beta}u^{\gamma}u^{\delta}+\:...\:,\label{ModuleX} \ee
  since in the normal coordinates $\partial_{\alpha}\partial_{\beta}X^{\mu}$ coincides with the second
  fundamental form $\Omega_{\alpha\beta}^{\mu}$ defined in (\ref{SFF}).

  We come back now to the calculation of $I$ in eq. (\ref{I1}). We are interested essentially in the pole of $I$, as explained
  above. Such a pole arises when the coordinates $\xi_1$ and $\xi_2$ come close to each other. Therefore we
  may take the internal integral in (\ref{I1}) just over a small neighborhood of the point $\xi_1$. The particular form
  of this neighborhood is unimportant, so we take it to be the circle of some small ``geodesic radius" $r$. We denote this circle
  by $C_r$. Later we will impose some constraints on $r$, but the final answer will be independent of $r$.

  Consider the internal integral in (\ref{I1})
    $$J(\xi_1)=\int\limits_{C_r}d^{\frac{D-2}{2}}\xi_2\sqrt{g}\Delta(\xi_1,\xi_2).$$
  We introduce the normal coordinates at $\xi_1$ both in $M$ and in $N$. Using (\ref{ModuleX}) we write the propagator
  in these coordinates
    $$\Delta(u,0)=\alpha\frac{1-\frac{D-2}{12}\tilde P_{\alpha\beta}u^{\alpha}u^{\beta}+\:...} {\Bigl(u^2-\frac{1}{12}
    \delta_{\mu\nu}\Omega_{\alpha\beta}^{\,\mu}\Omega_{\gamma\delta}^{\,\nu}u^{\alpha}u^{\beta}u^{\gamma}u^{\delta}+
    \:...\Bigr)^{\frac{D-2}{2}}}.$$ The notation here and in what follows reproduces that of section 3: a tilde
    denotes a pull-back of a corresponding tensor from $M$ to $N$,
    all internal quantities are denoted by a hat.
  The expansion of $\sqrt{g}$ in normal coordinates is
    $$\sqrt{g}=1+\frac16\hat R_{\alpha\beta}u^{\alpha}u^{\beta}+\:...$$
  In our coordinate system $J$ is taken at the origin and is given by
    \be J(0)=\alpha\int\limits_{C_r}\frac{1+\frac16\hat R_{\alpha\beta}u^{\alpha}u^{\beta}-\frac{D-2}{12}\tilde
       P_{\alpha\beta}u^{\alpha}u^{\beta}+\:...\:}{(u^2)^{\frac{D-2}{2}}\Bigl(1-\frac{1}{12u^2}\delta_{\mu\nu}
       \Omega_{\alpha\beta}^{\,\mu}\Omega_{\gamma\delta}^{\,\nu}u^{\alpha}u^{\beta}u^{\gamma}u^{\delta+\:...\:}
       \Bigr)^{\frac{D-2}{2}}}\,d^{\frac{D-2}{2}}u.\label{J} \ee
  From this expression we see that $r$ should be small enough so that the denominator in the integrand wouldn't vanish.

  Now we expand the integrand in powers of $u$. Since the integration goes over a circle we can introduce polar
  coordinates. Then we can replace $u^{\alpha}u^{\beta}$ by $u^2\delta_{\alpha\beta}/d$ etc. and then turn back to the
  denominator of the form of (\ref{J}). The integral obtained in this way is not equal to $J$, but differs from it only by finite
  terms, so the difference is unimportant for us. We get
    \be
       J(0)=\alpha\int\limits_{C_r}\frac{(u^2)^{-\frac{D-2}{2}}\Bigl(1+\frac{\hat R}{3(D-2)}u^2-\frac16Pu^2+\Bigr)\,d^{
       \frac{D-2}{2}}u}{\Bigl[1-\frac{u^2}{3(D-2)(D+2)}\delta_{\mu\nu}\Omega_{\alpha\beta}^{\,\mu}\Omega_{\gamma\delta}^
       {\,\nu}(\delta_{\alpha\beta}\delta_{\gamma\delta}+\delta_{\alpha\gamma}\delta_{\beta\delta}+\delta_{\alpha\delta}
       \delta_{\beta\gamma})\Bigr]^{\frac{D-2}{2}}}\,+finite,
    \ee
  where $P=g^{\alpha\beta}\tilde P_{\alpha\beta}$. We use the following notation
    \begin{eqnarray}
       K&=&\frac{\hat R}{3(D-2)}-\frac{P}{6},\\
       L&=&\frac{1}{3(D-2)(D+2)}\,\delta_{\mu\nu}\Omega_{\alpha\beta}^{\,\mu}\Omega_{\gamma\delta}^{\,\nu}
       \,(\delta_{\alpha\beta}\delta_{\gamma\delta}+\delta_{\alpha\gamma}\delta_{\beta\delta}+\delta_{\alpha\delta}
       \delta_{\beta\gamma})\label{L}.
    \end{eqnarray}
  The angular integration is easily performed. The result is
    \be J(0)=\alpha\,\sigma_{\frac{D-2}{2}}\int\limits_{0}^{r}\frac{(1+K\,u^2)\,u^{\frac{D-2}{2}-1}}{(u^2)^{\frac{D-2}{2}}
       (1-L\,u^2)^{\frac{D-2}{2}}}du,\ee
  where like in the eq. (\ref{al}) $\sigma_d$ means the area of a unit sphere in the $d$-dimensional Euclidean space.
  By changing the integration variable the last integral can be transformed to the form
    $$J(0)=\alpha\,\sigma_{\frac{D-2}{2}}\,r^{1-D/2}\int\limits_{0}^{1}\frac{(t^{-D/4-1/2}+Kt^{-D/4+1/2})}{(1-L\,r^2\,t)^
    {\frac{D-2}{2}}}\,dt.$$
  This integral can be evaluated using the following identity:
    \be \int\limits_{0}^{1}t^{b-1}(1-t)^{c-b-1}(1-zt)^{-a}dt=\frac{\Gamma(b)\Gamma(c-b)}{\Gamma(c)}\,\,{}_2F_1(a,b,c;z), \ee
  where ${}_2F_1(a,b,c;z)$ is a hypergeometric function defined as
    \be {}_2F_1(a,b,c;z)=\frac{\Gamma(c)}{\Gamma(a)\Gamma(b)}\sum\limits_{n=o}^{\infty}\frac{\Gamma(a+n)\Gamma(b+n)}
       {\Gamma(c+n)}\frac{z^n}{n!}. \ee
  The use of these identities gives
    \be  J(0)=r^{1-D/2}\frac{\alpha\,\sigma_{\frac{D-2}{2}}
       }{\Gamma(\frac{D-2}{2})}\sum\limits_{n=0}^{\infty}\Gamma\Bigl(
       \frac{D-2}{2}+n\Bigr)\frac{(r^2L)^n}{n!}\,\Biggl[\frac{1}{-\frac{D}{4}+\frac{1}{2}+n}+\frac{r^2K}{-\frac{D}{4}+\frac{3}{2}+n}\Bigg].\ee
  $\Gamma$-functions don't produce poles, so the only poles are those that come from the denominators in brackets. At
  $D\approx6$ there are poles at $n=0,1$. So the divergent part of $J$ is
    \be J(0)=\alpha\,\sigma_{\frac{D-2}{2}}r^{3-D/2}\frac{1}{\frac32-\frac{D}{4}}\Bigl(K+\frac{D-2}{2}L\Bigr).\ee
  We see that at $D=6$ $r$ disappears as we expected.

  We will now bring $L$ into the covariant form. We defined it in eq. (\ref{L}) as
    \be
    L=\frac{1}{3(D-2)(D+2)}\,\delta_{\mu\nu}\Omega_{\alpha\beta}^{\,\mu}\Omega_{\gamma\delta}^{\,\nu}\,
       (\delta_{\alpha\beta}\delta_{\gamma\delta}+\delta_{\alpha\gamma}\delta_{\beta\delta}+\delta_{\alpha\delta}
       \delta_{\beta\gamma}). \label{L1}\ee
  There are 2 kinds of terms here. In order to calculate $\delta_{\mu\nu}\delta_{\alpha\beta}\delta_{\gamma\delta}
  \Omega_{\alpha\beta}^{\,\mu}\Omega_{\gamma\delta}^{\,\nu}$ we notice that in the normal coordinates $g^{\alpha\beta}
  \Omega_{\alpha\beta}^{\,\mu}\equiv\Box X^{\mu}$, where $\Box X^{\mu}$ is a mean curvature vector,
  and therefore this term is $\Box X^{\mu}\Box X_{\mu}$. In order to
  ``covariantize" other terms in (\ref{L1}) we use the Gauss-Codazzi equation
    \be \hat R_{\alpha\beta\gamma\delta}=\tilde R_{\alpha\beta\gamma\delta}+G_{\mu\nu}(\Omega_{\alpha\delta}^{\,\mu}
       \Omega_{\beta\gamma}^{\,\nu}-\Omega_{\alpha\gamma}^{\,\mu}\Omega_{\beta\delta}^{\,\nu}),\label{GaussCodazzi}\ee
  which can be quickly verified in the normal coordinates. Here, as usual, $\hat R_{\alpha\beta\gamma\delta}$ is the
  Riemann tensor of $N$ and $\tilde R_{\alpha\beta\gamma\delta}$ is the pull-back of the Riemann tensor of $M$ to $N$.
  Contracting the eq. (\ref{GaussCodazzi}) with $\delta_{\alpha\beta}\delta_{\gamma\delta}$, we get
    \be
       \delta_{\mu\nu}\delta_{\alpha\gamma}\delta_{\beta\delta}\Omega_{\alpha\beta}^{\,\mu}\Omega_{\gamma\delta}^{\,\nu}=
       \Box X^{\mu}\Box X_{\mu}+\hat R-\delta_{\alpha\gamma}\delta_{\beta\delta}\tilde C_{\alpha\beta\gamma\delta}-
       (D-4)P.
    \ee
  We denote $\delta_{\alpha\gamma}\delta_{\beta\delta}\tilde C_{\alpha\beta\gamma\delta}$ as $C$. Plugging all these
  into the eq. (\ref{L1}) gives
    $$L=\frac{1}{3(D-2)(D+2)}\Bigl(3\Box X^{\mu}\Box X_{\mu}+2\hat R-2C-4P\Bigr).$$
  Finally, the pole part of the internal integral is
    \begin{multline*}
       J(0)=2\,\alpha\,\sigma_{\frac{D-2}{2}}\,r^{3-D/2}\frac{1}{3-D/2}\Bigl[\frac{2D}{3(D-2)(D+2)}\,\hat R-\\
       -\frac{1}{3(D-2)}\,C+
       \frac{2}{D+2}\,\Box X^{\mu}\Box X_{\mu}-\Bigl(\frac16+\frac{2}{3(D+2)}\Bigr)P\Bigr].\hspace{1cm}
    \end{multline*}
  Everywhere except the pole we can put $D=6$. Then
    $$J(0)=2\,\alpha\,\sigma_2\frac{1}{3-D/2}\Bigl(\frac18\hat R-\frac{1}{24}C+\frac{1}{16}(\Box X^{\mu}\Box X_{\mu}-4P)
    \Bigl).$$
  We calculated $J$ in the normal coordinates. But the result is fully covariant and therefore  independent of any
  particular coordinate frame. Having calculated $J$, we can finally write down the pole part of $I$:
    \be I_{pole}=\frac{\sigma_2}{\sigma_6}\frac{1}{D-6}\int\limits_N\sqrt{g}\Bigl(\frac18\hat R-\frac{1}{24}C+\frac{1}{16}
       (\Box X^{\mu}\Box X_{\mu}-4P)\Bigr)d^{\,2}\xi.\label{FinalResult}\ee
  Here we substituted the precise expression for $\alpha$ from (\ref{al}).

  The conformal variations of various terms of the last expression are
    \begin{eqnarray*}
       \delta(\sqrt{g}\,\hat R)&=&\frac{D-6}{2}\sqrt{g}\,\hat R\,\sigma+(D-3)\sqrt{g}\,\Box\sigma,\\
       \delta(\sqrt{g}\,C)&=&\frac{D-6}{2}\sqrt{g}\,C\sigma,\\
       \delta(\sqrt{g}\,\Box X^{\mu}\Box X_{\mu})&=&\frac{D-6}{2}\sqrt{g}\,\Box X^{\mu}\Box X_{\mu}\,\sigma-
          (D-2)\sqrt{g}\,\Box X^{\mu}\partial_{\mu}\sigma,\\
       \delta(\sqrt{g}\,P)&=&\frac{D-6}{2}\sqrt{g}\,P\sigma+\sqrt{g}\,\Box\sigma-\sqrt{g}\,\Box X^{\mu }\partial_{\mu}
       \sigma.
    \end{eqnarray*}
  All the terms proportional to $\Box\sigma$ vanish when integrated over $N$. Therefore the conformal variation of the
  pole of $I$ is
    \be \delta I_{pole}=\frac{1}{8\pi^2}\int\limits_{N}d^2\xi\sqrt{g}\,\Bigl[\hat R\sigma-\frac{1}{3}C\sigma
    +\frac{1}{2}(\Box X^{\mu}\Box X_{\mu}-4P)\,\sigma-\Box X^{\mu}\partial_{\mu}\sigma\Bigr].\ee
  The pole cancels as expected. This expression reproduces the
  result of \cite{Gustavsson:2004gj} for the considered model and is in a complete agreement with the
  results of \cite{Henningson:1999xi} discussed in the previous part (in a sense that there appear the same terms in the expression for the
  anomaly). However, as discussed after eq. (\ref{HSResult}) the
  last term represents a trivial anomaly, and therefore only 3
  first terms are relevant: \be \delta I_{pole}=\frac{1}{8\pi^2}\int\limits_{N}d^2\xi\sqrt{g}\,\Bigl[\hat R\sigma-\frac{1}{3}C\sigma
    +\frac{1}{2}(\Box X^{\mu}\Box X_{\mu}-4P)\,\sigma\Bigr].\ee

  Now we can understand the types of various anomalies.
  The first term which is proportional to $\hat R\sigma$ appeared as the conformal variation of $\hat R$ and is of the
  type A (see the eqs. (\ref{WD})-(\ref{R2DS})). Two other terms are of the type B since they emerge as conformal
  variations of nonvanishing expressions.


  \section{Conclusion}
   ~~~~In this paper we have considered the general features of the GW anomaly. It apparently occurs
   in the spaces of dimension $D=2+4n,\quad n=0,1,2$ etc. We showed that in general the GW anomaly consists
   of 2 groups of terms: internal anomalies that are the usual trace anomalies on the submanifold and external anomalies that are
   due to the ambient space. We argued also that all the external anomalies were of type B.

   For the lowest possible dimension $6$ we found all possible expressions for the anomaly and investigated their types
   using the model of free massless scalar field coupled minimally to the gravity. We found that there were 2 external
   anomalies of type B and a single internal anomaly of type A.

   Using these results we can make a few further guesses about general features of the GW anomalies. It seems natural
   to guess that the internal anomalies will always behave as described in \cite{Deser:1993yx}, i.e. among them there will always be
   a single type A anomaly proportional to the Euler density and a few anomalies
   of type B. In addition there will be also a few external anomalies of type B.
   Then for the 10-dimensional case (the next non-trivial one) with the 4-dimensional submanifold $N$ we may
   guess that there are 2 internal anomalies, namely
     \begin{eqnarray*}
        \Delta_1^{int}&=&\int\limits_Nd^4\xi\sqrt{g}\,(\hat R_{\alpha\beta\gamma\delta}\hat R^{\alpha\beta\gamma\delta}
        -4\hat R_{\alpha\beta}\hat R^{\alpha\beta}+\hat R^2)\,\sigma,\\
        \Delta_2^{int}&=&\int\limits_Nd^4\xi\sqrt{g}\,\hat C_{\alpha\beta\gamma\delta}\hat C^{\alpha\beta\gamma\delta}\sigma
     \end{eqnarray*}
   in the obvious notation. Here $\Delta_1$ is proportional to the Euler density and hence is of type A and $\Delta_2$
   is of type B. In addition we expect a few additional external
   terms in the anomaly. Our prediction is that all these are of
   type B. It is interesting to check this prediction by an explicit
   calculation.

\section*{Acknowledgements}

I am grateful to A. Schwimmer for discussions.

\end{document}